\documentclass[%
reprint,
superscriptaddress,
amsmath,amssymb,
aps,
prl,
]{revtex4-2}

\usepackage{blindtext}
\usepackage{graphicx}
\usepackage{dcolumn}
\usepackage{bm}
\usepackage[colorlinks=true,
            linkcolor=black,
            citecolor=cyan,
            urlcolor=black]{hyperref}
\usepackage[mathlines]{lineno}
\usepackage{lipsum}
\usepackage{booktabs}
\usepackage{layouts}
\usepackage{multirow}
\usepackage{xcolor}  
\usepackage{soul}
\usepackage{orcidlink}
\usepackage{textgreek}
\usepackage{braket}
\usepackage{siunitx}
\usepackage[normalem]{ulem}
\usepackage[markup]{changes}

\definechangesauthor[name={Author},color=blue]{au}
\setaddedmarkup{\textcolor{green!60!black}{#1}}
\setdeletedmarkup{\textcolor{red!100!black}{\sout{#1}}}
\definecolor{mplC1}{RGB}{255,127,14}
\setcommentmarkup{\textcolor{red}{#1}}

\setauthormarkup{}

\makeatletter
\renewcommand{\fnum@figure}{FIG.~\thefigure}
\renewcommand{\fnum@table}{TABLE~\thetable}

\makeatother

\begin{document}


\title{
Bound-state QED test above the Schwinger limit with kaonic fluorine
}
%

\author{F. Clozza\orcidlink{0009-0002-3298-0624}}
\thanks{Contact author: \texttt{francesco.clozza@lnf.infn.it} (Corresponding Author)}
\affiliation{Laboratori Nazionali di Frascati INFN, Frascati, Italy}
\affiliation{Università degli studi di Roma Tor Vergata, Dipartimento di Fisica, Roma, Italy}

\author{S. Manti\orcidlink{0000-0003-3770-0863}}
\affiliation{Laboratori Nazionali di Frascati INFN, Frascati, Italy}

\author{F. Sgaramella\orcidlink{0000-0002-0011-8864}}
\affiliation{Laboratori Nazionali di Frascati INFN, Frascati, Italy}

\author{L. Abbene\orcidlink{0000-0001-9633-6606}}
\affiliation{Department of Physics and Chemistry (DiFC), Emilio Segrè, University of Palermo, Palermo, Italy}
\affiliation{Laboratori Nazionali di Frascati INFN, Frascati, Italy}


\author{F. Artibani\orcidlink{0009-0000-8905-3165}}
\affiliation{Laboratori Nazionali di Frascati INFN, Frascati, Italy}
\affiliation{Università degli studi di Roma Tre, Dipartimento di Fisica, Roma, Italy}

\author{M. Bazzi\orcidlink{0000-0002-1699-7138}}
\affiliation{Laboratori Nazionali di Frascati INFN, Frascati, Italy}

\author{G. Borghi\orcidlink{0000-0001-8488-4728}}
\affiliation{Politecnico di Milano, Dipartimento di Elettronica, Informazione e Bioingegneria, Milano, Italy}
\affiliation{INFN Sezione di Milano, Milano, Italy}

\author{D. Bosnar\orcidlink{0000-0003-4784-393X}}
\affiliation{Department of Physics, Faculty of Science, University of Zagreb, Zagreb, Croatia}

\author{M. Bragadireanu\orcidlink{0009-0001-5217-6003}}
\affiliation{IFIN-HH, Institutul National pentru Fizica si Inginerie Nucleara Horia Hulubei, 30 Reactorului, 077125, Magurele, Romania}

\author{A. Buttacavoli\orcidlink{0000-0002-7188-3651}}
\affiliation{Department of Physics and Chemistry (DiFC), Emilio Segrè, University of Palermo, Palermo, Italy}
\affiliation{Laboratori Nazionali di Frascati INFN, Frascati, Italy}

\author{M. Carminati\orcidlink{0000-0001-9734-3007}}
\affiliation{Politecnico di Milano, Dipartimento di Elettronica, Informazione e Bioingegneria, Milano, Italy}
\affiliation{INFN Sezione di Milano, Milano, Italy}

\author{A. Clozza\orcidlink{0000-0003-2133-1725}}
\affiliation{Laboratori Nazionali di Frascati INFN, Frascati, Italy}

\author{L. De Paolis\orcidlink{0000-0002-4203-9902}}
\affiliation{Laboratori Nazionali di Frascati INFN, Frascati, Italy}

\author{R. Del Grande\orcidlink{0000-0002-7599-2716}}
\affiliation{Faculty of Nuclear Sciences and Physical Engineering, Czech Technical University in Prague, Břehovà 7, 115 19, Prague, Czech Republic}
\affiliation{Laboratori Nazionali di Frascati INFN, Frascati, Italy}

\author{K. Dulski\orcidlink{0000-0002-4093-8162}}
\affiliation{Laboratori Nazionali di Frascati INFN, Frascati, Italy}
\affiliation{Faculty of Physics, Astronomy, and Applied Computer Science, Jagiellonian University, Kraków, Poland}
\affiliation{Center for Theranostics, Jagiellonian University, Krakow, Poland}


\author{C. Fiorini\orcidlink{0000-0002-1157-0143}}
\affiliation{Politecnico di Milano, Dipartimento di Elettronica, Informazione e Bioingegneria, Milano, Italy}
\affiliation{INFN Sezione di Milano, Milano, Italy}

\author{I. Friščić\orcidlink{0000-0002-4743-0572}}
\affiliation{Department of Physics, Faculty of Science, University of Zagreb, Zagreb, Croatia}

\author{C. Guaraldo\orcidlink{0000-0002-8923-3438}}
\thanks{Deceased}
\affiliation{Laboratori Nazionali di Frascati INFN, Frascati, Italy}

\author{M. A. Iliescu\orcidlink{0009-0003-3859-5679}}
\affiliation{Laboratori Nazionali di Frascati INFN, Frascati, Italy}

\author{P. Indelicato\orcidlink{0000-0003-4668-8958}}
\affiliation{Laboratoire Kastler Brossel, Sorbonne Université, CNRS, ENS-PSL Research University, Collège de France, Case 74; 4, place Jussieu, F-75005 Paris, France}

\author{M. Iwasaki\orcidlink{0000-0002-3460-9469}}
\affiliation{RIKEN, Tokyo, Japan}

\author{A. Khreptak\orcidlink{0000-0002-9482-9770}}
\affiliation{Faculty of Physics, Astronomy, and Applied Computer Science, Jagiellonian University, Kraków, Poland}
\affiliation{Center for Theranostics, Jagiellonian University, Krakow, Poland}
\affiliation{Laboratori Nazionali di Frascati INFN, Frascati, Italy}

\author{J. Marton\orcidlink{0009-0003-1912-285X}}
\thanks{Now at Atominstitut, Technical University Vienna, Stadionallee 2, 1020 Vienna, Austria}
\affiliation{Stefan Meyer Institute for Subatomic Physics, Vienna, Austria}

\author{P. Moskal\orcidlink{0000-0001-5644-5963}}
\affiliation{Faculty of Physics, Astronomy, and Applied Computer Science, Jagiellonian University, Kraków, Poland}
\affiliation{Center for Theranostics, Jagiellonian University, Krakow, Poland}


\author{H. Ohnishi\orcidlink{0000-0002-0615-3214}}
\affiliation{Research Center for Accelerator and Radioisotope Science (RARiS), Tohoku University, Sendai, Japan}

\author{K. Piscicchia\orcidlink{0000-0001-6879-452X}}
\affiliation{Centro Ricerche Enrico Fermi, Museo Storico della Fisica e Centro Studi e Ricerche "Enrico Fermi", Roma, Italy}
\affiliation{Laboratori Nazionali di Frascati INFN, Frascati, Italy}

\author{F. Principato\orcidlink{0000-0003-2787-0877}}
\affiliation{Department of Physics and Chemistry (DiFC), Emilio Segrè, University of Palermo, Palermo, Italy}
\affiliation{Laboratori Nazionali di Frascati INFN, Frascati, Italy}

\author{A. Scordo\orcidlink{0000-0002-7703-7050}}
\affiliation{Laboratori Nazionali di Frascati INFN, Frascati, Italy}

\author{M. Silarski\orcidlink{0000-0003-2206-0963}}
\affiliation{Faculty of Physics, Astronomy, and Applied Computer Science, Jagiellonian University, Kraków, Poland}

\author{D. Sirghi\orcidlink{0009-0002-7486-025X}}
\affiliation{Centro Ricerche Enrico Fermi, Museo Storico della Fisica e Centro Studi e Ricerche "Enrico Fermi", Roma, Italy}
\affiliation{Laboratori Nazionali di Frascati INFN, Frascati, Italy}
\affiliation{IFIN-HH, Institutul National pentru Fizica si Inginerie Nucleara Horia Hulubei, 30 Reactorului, 077125, Magurele, Romania}

\author{F. Sirghi\orcidlink{0000-0002-6143-3200}}
\affiliation{Laboratori Nazionali di Frascati INFN, Frascati, Italy}
\affiliation{IFIN-HH, Institutul National pentru Fizica si Inginerie Nucleara Horia Hulubei, 30 Reactorului, 077125, Magurele, Romania}

\author{M. Skurzok\orcidlink{0000-0002-4794-5154}}
\affiliation{Faculty of Physics, Astronomy, and Applied Computer Science, Jagiellonian University, Kraków, Poland}
\affiliation{Center for Theranostics, Jagiellonian University, Krakow, Poland}
\affiliation{Laboratori Nazionali di Frascati INFN, Frascati, Italy}

\author{J. Sommerfeldt\orcidlink{0000-0002-3471-7494}}
\affiliation{Laboratoire Kastler Brossel, Sorbonne Université, CNRS, ENS-PSL Research University, Collège de France, Case 74; 4, place Jussieu, F-75005 Paris, France}

\author{A. Spallone\orcidlink{0009-0000-2111-8014}}
\affiliation{Laboratori Nazionali di Frascati INFN, Frascati, Italy}

\author{K. Toho\orcidlink{0009-0001-3245-1418}}
\affiliation{Research Center for Accelerator and Radioisotope Science (RARiS), Tohoku University, Sendai, Japan}
\affiliation{Laboratori Nazionali di Frascati INFN, Frascati, Italy}


\author{O. Vazquez Doce\orcidlink{0000-0001-6459-8134}}
\affiliation{Laboratori Nazionali di Frascati INFN, Frascati, Italy}

\author{J. Zmeskal\orcidlink{0000-0003-0815-0639}}
\thanks{Deceased}
\affiliation{Stefan Meyer Institute for Subatomic Physics, Vienna, Austria}

\author{C. Curceanu\orcidlink{0000-0002-1990-0127}}
\affiliation{Laboratori Nazionali di Frascati INFN, Frascati, Italy}
\affiliation{IFIN-HH, Institutul National pentru Fizica si Inginerie Nucleara Horia Hulubei, 30 Reactorului, 077125, Magurele, Romania}

\collaboration{SIDDHARTA-2 Collaboration}

\date{\today}

\begin{abstract}
Kaonic atoms, formed when a negatively charged kaon replaces an electron in an atomic orbit, provide access to bound-state quantum electrodynamics (BSQED) in electromagnetic fields far stronger than in ordinary atoms. Here, we report an experimental test of BSQED in a regime where the mean Coulomb field exceeds the Schwinger limit. Using high-precision x-ray spectroscopy of kaonic fluorine with the SIDDHARTA-2 experiment at DA$\Phi$NE, corresponding to an integrated luminosity of 22.4~pb$^{-1}$, we observe transitions involving the 4f and 3d levels, probing field-to-Schwinger-limit ratios of 1.11 and 3.70, respectively. The measured transition energies agree with state-of-the-art Dirac--Fock calculations. In particular, the 5g--4f transition showing a residual of 5.8\,$\pm$\,4.7\,(stat.)\,$\pm$\,5.5\,(syst.)~eV and a $\sim$\,9$\sigma$ sensitivity to QED contributions. These results provide a direct test of BSQED in the strong-field regime of QED above the Schwinger limit, opening a new avenue for precision studies in extreme electromagnetic fields.
\end{abstract}

\maketitle
%
\newif\ifdraft
\draftfalse
%
\ifdraft
\section{Introduction}
\fi
Exotic atoms, atomic systems in which a negatively charged particle (\textit{e.g.} $\mu^-$, $K^-$, $\bar{p}$, etc.) replaces an electron, are powerful tools to probe fundamental interactions \cite{Curceanu:2026zjg}. In this framework, kaonic atoms have been successfully used to investigate strong interaction in the low-energy regime \cite{SIDDHARTA:2011dsy,J-PARCE62:2022qnt,Curceanu:2019uph,Curceanu:2026zjg} to extract hadronic scattering lengths and to test effective field theories at threshold \cite{Curceanu:2026zjg,Obertova:2022des,Shevchenko:2016wnu}. The scope of kaonic atoms spectroscopy extends beyond hadronic physics, as they also offer a unique environment for testing quantum electrodynamics (QED) \cite{Paul:2020cnx}. QED is the most precisely tested theory among fundamental interactions; in its bound-state formulation (BSQED), theoretical calculations reach parts-per-billion precision, as demonstrated with the hydrogen 2s--1s transition \cite{Parthey:2011lfa}. For intermediate and high atomic numbers $Z$, however, the perturbative expansion in ($\alpha Z$), with $\alpha$ the fine-structure constant, converges slowly, limiting the achievable theoretical precision. Highly charged ions (HCI) \cite{Indelicato:2019nij} have traditionally been used for testing BSQED in strong electric fields \cite{ullmann2017high, Loetzsch:2024pae, Morgner:2023ghx} in intermediate- and high-$Z$ systems. The achievable precision of such tests is nonetheless limited by nuclear-structure effects -- such as the finite nuclear size (FNS) and nuclear polarization -- which become comparable to higher-order QED contributions for the intermediate- and high-$Z$ ions \cite{Yerokhin:2003pq, Volotka:2013vpa}. In this context, exotic atoms have emerged as a competitive alternative to HCI \cite{Paul:2020cnx} and they have been already proven to be valuable systems for investigating BSQED \cite{PhysRevLett.130.173001}. In particular, the SIDDHARTA-2 Collaboration at the DA$\Phi$NE collider of the  National Laboratories of Frascati (INFN-LNF) recently demonstrated that kaonic atoms are a key alternative to muonic and antiprotonic ones and a new frontier for BSQED tests, by reporting the first precision x-ray spectroscopy of kaonic neon \cite{Sgaramella:2024klx, SIDDHARTA-2:2025mwc}. In kaonic atoms, the compact Bohr orbits further enhance the electric field experienced by the kaon by a factor of $\mathcal{O}(\mu^2/m_e^2)$, where $\mu$ denotes the reduced mass of the kaon and $m_e$ the mass of the electron. This makes kaonic atoms natural laboratories for investigating strong-field QED (SFQED), \textit{i.e.}~QED in the presence of intense electromagnetic fields. In this regime, vacuum polarization and higher-order radiative corrections are strongly enhanced \cite{Wichmann:1956zz, Hattori:2023egw, Dunne:2004nc}, leading to measurable eV-scale effects in keV transitions. Certain transitions in exotic atoms probe electric fields approaching or exceeding the Schwinger critical value $\mathcal{E}_c = m_e^2c^3/(e\hbar) \simeq 1.3 \times 10^{18}$~V/m \cite{Schwinger:1951nm}. By approaching this threshold, QED enters a non-linear domain in which higher-order contributions become increasingly important and the perturbative expansion converges more slowly \cite{Hattori:2023egw, Dunne:2004nc}. This results in enhanced corrections to the binding energies in exotic atoms, particularly for states closer to the nucleus. Such supercritical fields amplify non-linear QED phenomena, including vacuum birefringence, photon-photon scattering, and higher-order radiative effects -- compactly described by the Heisenberg-Euler effective action \cite{Heisenberg:1936nmg} -- thus extending the experimental reach to SFQED and offering new avenues to search for physics beyond the standard model \cite{Liu:2025ows}. Electromagnetic fields exceeding the Schwinger limit can also be found in extreme astrophysical environments such as magnetars and black holes \cite{Harding:2006qn,Rigoselli:2024nyk,Wondrak:2023zdi}, making laboratory-scale probes of critical fields an essential benchmark for interpreting such phenomena. Previous studies of muonic, pionic, and antiprotonic atoms accessed electric fields exceeding the Schwinger limit by measuring low-$n$ transitions in high-$Z$ systems, such as 3d-2p in $\pi$Ti \cite{Powers:1980ie}, 5g-4f in $\mu$Pb \cite{Dubler:1978rq} and 14n-13m in $\bar{p}$Xe \cite{gotta2008x}. However, in these systems, contributions from electron screening, FNS, or strong-interaction effects are often non-negligible and can reach the eV to tens-of-eV scale, thereby limiting a direct and stringent comparison with BSQED predictions in the strong-field regime.
The accurate interpretation of these effects requires theoretical models capable of including both relativistic and radiative contributions. Multiconfiguration Dirac-Fock (MCDF) methods, extended to account for QED corrections (Uehling, Wichmann-Kroll, self-energy) and FNS effects, provide sub-eV accuracy and are crucial for disentangling QED effects from hadronic and nuclear contributions. Comparison of measured low-$n$ transition energies with all-order MCDF calculations \cite{sommerfeld2026allorder} directly tests the internal consistency of QED in a regime where its perturbative description is challenged. Moreover, a precise comparison between experiment and QED predictions in this regime enables stringent searches for physics beyond the standard model, particularly for new interaction mediators in the 0.1--10~MeV mass range \cite{Liu:2025ows}. Driven by these developments, the exploration of SFQED  has gained renewed theoretical and experimental attention in recent years \cite{Fedotov:2022ely, Sarri:2025qng, Okumura:2024llw, Baptista:2025kai, Schulthess:2026pmg}. \noindent
In this Letter we report a high-precision measurement of x-ray transitions in electric fields exceeding the Schwinger critical one using kaonic fluorine (KF), performed by the SIDDHARTA-2 Collaboration. The measured transition energies are compared with state-of-the-art MCDF calculations incorporating all-order QED contributions, providing a stringent and direct test of BSQED in the strong-field regime.
%
\ifdraft
\section{Methods}
\fi

The KF measurement was performed in the summer of 2024 at the DA$\Phi$NE collider at INFN-LNF \cite{Milardi:2018sih,Milardi:2021khj,Milardi:2024efr} using the SIDDHARTA-2 setup \cite{Sirghi:2023wok}, for a total integrated luminosity of 22.4~pb$^{-1}$, collected over four days of data taking. KF x-ray lines were measured with 384 silicon drift detectors (SDDs), for a total sensitive area of 246~cm$^2$, with an energy resolution (FWHM) of 160~eV at 6.4~keV and a timing resolution of 450~ns \cite{Miliucci:2021wbj}. A 1-mm-thick Teflon (C$_2$F$_4$) foil served as fluorine target material. Dedicated calibration runs of the SDDs were performed using x-ray reference lines, based on a linear calibration function among the Bi L$\alpha$ line and the K$\alpha$ lines of Pd and Ag. Systematic uncertainties on the KF fluorescence lines were evaluated from calibration residuals of additional reference lines, Ba K$\alpha_1$ and Tm K$\alpha_1$. More details about the calibration procedure and related systematics are reported in Ref. \cite{Clozza:2026idt}. Other potential sources of systematic uncertainty \cite{Sgaramella:2024klx} were not included, as they are subleading with respect to the dominant energy calibration uncertainty. To extract transition energies of KF lines, a chi-squared fit of the x-ray spectrum was performed. The spectral response of the SDDs was modeled, for each peak, with a Gaussian function with a resolution parametrized in terms of Fano factor and electric noise of the SDDs \cite{Miliucci:2021wbj}, and within the acquired statistics, no asymmetry in the peaks was observed. The background was best described by a decreasing exponential function plus a constant offset. The fit to the spectrum was conducted covering an energy range from 9.8~keV to 54~keV.
Transition energies were calculated with the Multiconfiguration Dirac–Fock General Matrix Element (\textsc{mcdfgme}) code \cite{desclaux1975,indelicato1990} (version 2025.1), using the 2018 \textsc{codata} constants \cite{Tiesinga:2021myr}. Transitions in KF have a hyperfine splitting (HFS) \cite{Trassinelli:2006xs} which is negligible for the energy resolutions of this study, therefore only transitions between circular states were considered, with residual HFS effects included in the theoretical uncertainty budget. The leading vacuum-polarization contribution was treated self-consistently, while higher-order terms were included perturbatively. Separate all-order vacuum-polarization calculations \cite{Soff:1988zz,sommerfeld2026allorder} were performed as a consistency check and were found to give negligible corrections. Nuclear recoil contributions were also incorporated in the total energies following the method of Ref. \cite{Trassinelli:2006xs}. For the KF 4f--3d transition, a strong interaction shift $\epsilon_{3d}$ was included in the reported transition energy, calculated using a kaon–multinucleon scattering approach as described in Refs. \cite{Obertova:2025rso,Obertova:2022des}. Strong-interaction effects on states above the 3d level are negligible, and no induced shift was therefore included for these levels.
The theoretical uncertainty is dominated by nuclear effects and external input parameters, namely electronic screening and the kaon mass. For the nuclear contribution, a Fermi charge distribution was adopted for the nuclear density, and the associated uncertainty was estimated from the FNS correction by comparison with the point-nucleus approximation. The screening contribution, due to residual electrons left during the cascade \cite{Trassinelli:2016brc,kirch1999muonic}, was assessed by recalculating the transition energies for a configuration including the kaon and a single electron in the fluorine 1s orbital, and taking the resulting energy shift as an uncertainty estimate. The uncertainty from the kaon mass was estimated by varying it within its PDG $\pm 13$~keV uncertainty around the central value $M_{K^-} = 493.677$~MeV \cite{ParticleDataGroup:2024cfk}. The effect of unresolved hyperfine structure was estimated conservatively by taking half the full energy span of the allowed hyperfine components as an additional uncertainty. The FNS term produces a positive shift, electronic screening terms induce negative shifts, and the kaon-mass and HFS contributions are symmetric. Upward and downward theoretical uncertainties were therefore evaluated separately, resulting in an asymmetric uncertainty band used in the transition energy residuals. The total theoretical uncertainty was obtained by combining in quadrature the individual contributions, while accounting for the sign of the corresponding energy shifts.
Average electric field values experienced by the kaon were computed as radial averages of the electric field \cite{Paul:2020cnx}, weighted by the large and small components of the kaon wavefunction for the given level.
%
\ifdraft
\section{Results and Discussions}
\fi
%
\begin{figure*}[t]
        \centering \includegraphics[width=0.8\textwidth]{figures/1_Spectrum.png}
        \caption{KF x-ray spectrum and fit, showing experimental data (black), global fit (red), KF lines (pink), kaonic carbon (KC) and contaminant Bi and Ag lines (blue) and background (green). The pull plot (bottom panel) displays the normalized residuals of the fit.}
        \label{fig:KF_spectrum}
\end{figure*}
%

\begin{figure}[t]
    \centering
    \includegraphics[width=1.\linewidth]{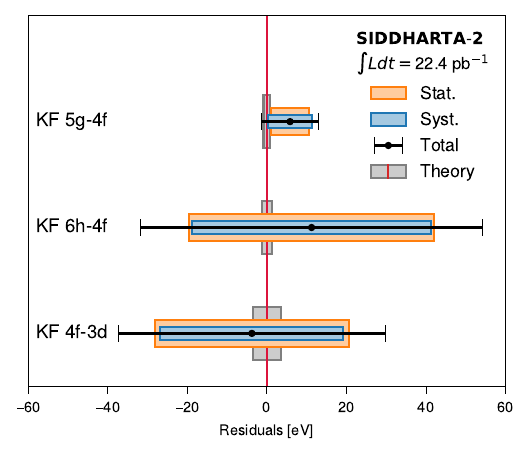}
    \caption{Energy residuals $E_{if}^{\text{(exp.)}} - E_{if}^{\text{(calc.)}}$ for the lowest-$n$ KF measured transitions. Error bars show statistical (blue), systematic (orange), and total experimental uncertainties. The shaded gray band indicates the theoretical uncertainty of the \textsc{mcdfgme} prediction arising from FNS and HFS effects, the kaon-mass input and electronic screening.}
    \label{fig:residuals}
\end{figure}

The x-ray spectrum of KF and the corresponding fit are shown in Fig. \ref{fig:KF_spectrum}. Several KF circular transitions are observed in the 10--50~keV range, with the most intense lines corresponding to the 6h--5g and 5g--4f transitions at 12.7~keV and 23.4~keV, respectively, while the 4f--3d transition is at 50.6~keV. The latter is described by a Gaussian profile in the fit as the intrinsic strong-interaction-induced width -- expected to be $\sim$20~eV from theoretical predictions \cite{Obertova:2025rso,Obertova:2022des} -- is much smaller than detector resolution (418\,$\pm$\,28~eV FWHM) at that energy \cite{Clozza:2026idt}, hence cannot experimentally be resolved. The HFS contributions are likewise negligible at this scale (see Table \ref{tab:KF_tab_SM} \cite{supplemental}), justifying the use of a single Gaussian component for all KF transitions. Additional weaker KF transitions with $\Delta n>1$ are also present in the spectrum. Alongside the KF lines, the kaonic carbon 5g--4f transition at 10.2~keV is observed, originating from kaons stopped inside the setup materials. Fluorescence lines of Bi and Ag are present, arising from the SDDs ceramic material, in agreement with our previous measurements \cite{Sgaramella:2023orc}.
Transition energies extracted from the fit are summarized in Table \ref{tab:KF_tab} for the six circular KF transitions. The table lists the measured energies $E_{if}^{\text{(exp.)}}$ with their statistical uncertainties $\delta E_{if}^{\text{(stat.)}}$, obtained from the fit, and systematic uncertainties $\delta E_{if}^{\text{(sys.)}}$, determined by the energy calibration. The most intense lines, the 6h--5g and 5g--4f transitions, are dominated by systematic uncertainties ($\sim$5$\div$7~eV), due to their large statistics, while all other lines are dominated by the statistical uncertainties ($\sim$20$\div$40~eV).
Table \ref{tab:KF_tab} also compares the measured transition energies with the corresponding \textsc{mcdfgme} calculations, reporting the associated theoretical uncertainties and the total QED contribution to the transition energy.
In Table \ref{tab:KF_tab_SM} of the supplemental material \cite{supplemental} the decomposition of these contributions into first- and second-order terms is reported. For the 4f--3d transition, the theoretical value additionally includes a negative shift $\epsilon_{3d}$ of -2~eV due to the strong-interaction effect on the 3d state \cite{Obertova:2025rso,Obertova:2022des}. 
The QED correction for this transition amounts to 222.99~eV, corresponding to 0.44\% of its total transition energy. This fractional contribution is more than twice that observed in kaonic neon \cite{SIDDHARTA-2:2025mwc}. This enhancement reflects the stronger mean Coulomb field experienced by the kaon in lower-$n$ states. Since QED contributions scale non-linearly with the local electric field strength rather than with the emitted photon energy, the observed relative increase provides direct evidence of the transition into a stronger-field bound-state regime.
Putting this result in the broader context of exotic atoms, it is clear that kaonic atoms provide a particularly advantageous platform for BSQED tests in strong fields. In muonic atoms, the fractional QED contribution typically remains at the $\times10^{-4}$ level, for example, about 0.04\% in muonic neon \cite{PhysRevLett.130.173001}, an order of magnitude smaller than observed here in KF. Antiprotonic atoms probe even stronger Coulomb fields \cite{Paul:2020cnx}, but FNS effects rapidly become dominant for low-$n$ states, limiting the clean isolation of QED terms.
Given the sizable magnitude of the QED contributions, theoretical uncertainties were evaluated to quantify the degree of agreement with the experimental results. A detailed breakdown of the individual uncertainty contributions -- FNS, electronic screening, kaon-mass term and HFS -- is provided in Table \ref{tab:KF_tab_SM} of the supplemental material \cite{supplemental}. The dominant theoretical uncertainties arise from the kaon mass \cite{SIDDHARTA-2:2025mwc}, reflecting its PDG uncertainty \cite{ParticleDataGroup:2024cfk}, and the HFS contribution, while the FNS and electron screening terms remain subleading at the sub-eV scale. The latter is included as an uncertainty estimate but is expected to be negligible for low-$n$ transitions, since the kaon orbits lie well within the electronic cloud, with radii about $\sim$50 times smaller than that of the fluorine 1s electronic orbital.
Within the combined experimental uncertainties, the measured energies are consistent with the calculated values, which motivates presenting the comparison in terms of residuals for the transitions involving the 4f level, as done in Fig. \ref{fig:residuals}, where FNS, HFS, kaon-mass, and screening components define the theoretical uncertainty band.
\begin{table}[t]
    \renewcommand{\arraystretch}{1.2}
    \centering
    \caption{Measured and calculated transition energies for KF. Reported are the experimental energies $E_{if}^{(\mathrm{exp.})} \pm (\mathrm{stat.}) \pm (\mathrm{sys.})$, the calculated values $E_{if}^{(\mathrm{calc.})}\,{}^{+\delta_{if}^{+}}_{-\delta_{if}^{-}}$, and the QED contributions $E_{if}^{(\mathrm{QED})}$. The theoretical uncertainties include FNS, kaon-mass, screening, and HFS effects. All values are given in eV.}
    \begin{ruledtabular}
    \begin{tabular}{@{}lccc@{}}
    \addlinespace[1mm]
    Line & $E_{if}^{(\mathrm{exp.})} \pm (\mathrm{stat.}) \pm (\mathrm{sys.})$ & $E_{if}^{(\mathrm{calc.})}\,{}^{+\delta_{if}^{+}}_{-\delta_{if}^{-}}$ & $E_{if}^{(\mathrm{QED})}$ \\
    \addlinespace[1mm]
    \hline
    \addlinespace[1mm]
6h--5g & $12673.9 \pm 5.9 \pm 7.1$ & $12684.69_{-0.43}^{+0.42}$ & $23.49$ \\
7i--5g & $20298.8 \pm 16.3 \pm 6.0$ & $20327.49_{-0.64}^{+0.59}$ & $32.26$ \\
5g--4f & $23383.2 \pm 4.7 \pm 5.5$ & $23377.38_{-1.06}^{+1.06}$ & $67.54$ \\
8k--5g & $25279.6 \pm 36.2 \pm 5.0$ & $25285.56_{-0.85}^{+0.70}$ & $35.65$ \\
6h--4f & $36073.3 \pm 30.7 \pm 30.0$ & $36062.07_{-1.28}^{+1.27}$ & $91.02$ \\
4f--3d & $50586.7 \pm 24.3 \pm 23.0$ & $50590.48_{-4.36}^{+4.36}$ & $222.99$ \\
    \end{tabular}
    \end{ruledtabular}
    \label{tab:KF_tab}
\end{table}

\begin{figure}[t]
    \centering
    \includegraphics[width=1.05\linewidth]{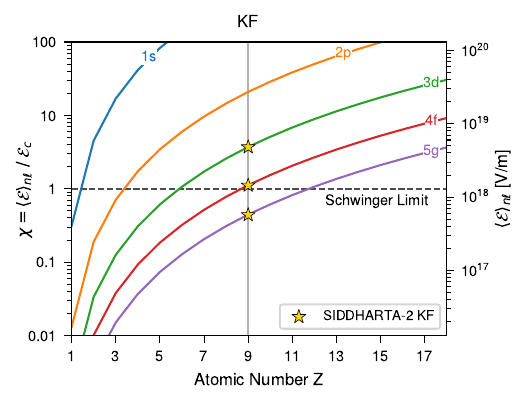}
    \caption{Field-to-Schwinger-limit ratio $\chi$\,=\,$\braket{\mathcal{E}}_{n\ell}/\mathcal{E}_c$ in kaonic atoms, as a function of the nuclear charge $Z$ for levels from 1s to 6h. The horizontal dashed line indicates the Schwinger critical field $\mathcal{E}_c = m_e^2c^3/(e\hbar) \simeq 1.3 \times 10^{18}$~V/m ($\chi$\,=\,1). The vertical marker highlights KF, for which the $n\ell$ states involved in the measured transitions probe fields at and above $\mathcal{E}_c$ ($\chi$\,$>$\,1). Starred points mark the corresponding $\chi$ values for the KF states accessed in this measurement.
}
    \label{fig:schwinger_lim}
\end{figure}

%
In Fig. \ref{fig:residuals} we compare the experimental transition energies with the corresponding calculated values for the KF transitions involving the lowest-$n$ levels. Both statistical and systematic uncertainties are shown, together with the total uncertainty obtained by summing them in quadrature. The theoretical uncertainties are also reported, with the individual contributions treated with their proper sign, reflecting their effect on the calculated energies. In this strong-field regime, we find agreement for all measured transitions with the theoretical predictions within the uncertainties. The most intense line, \textit{i.e.}~the 5g--4f transition, which is free from strong-interaction effects, shows an energy residual of 5.8\,$\pm$\,4.7\,(stat.)\,$\pm$\,5.5\,(syst.)~eV, corresponding to a $\sim$\,9$\sigma$ sensitivity to the QED contribution and demonstrating agreement between experiment and theory in the critical-field regime of BSQED.
%
To quantify the strong field sampled by these transitions, we evaluate the average electric field experienced by the kaon in the atomic orbit $\braket{\mathcal{E}}_{n\ell}$, a quantity commonly used to define the strong-field scale in exotic atoms \cite{Indelicato:2019nij}. In Table \ref{tab:efield_schwinger} we report the values of $\braket{\mathcal{E}}_{n\ell}$ in different orbital states $n\ell$ together with the ratio $\chi$\,=\,$\braket{\mathcal{E}}_{n\ell}/\mathcal{E}_c$ \cite{Fedotov:2022ely} and their mean orbital radii $\braket{r}_{n\ell}$. Remarkably, the threshold value $\chi$\,=\,1 is exceeded starting from the 4f level of KF, with a ratio of 1.11, and reaches on the 3d state a value of 3.70.
The observation of transitions probing electric fields above the Schwinger limit provides access to the critical-field regime of BSQED. In particular, the 5g--4f line enables a clean test of BSQED. Combined with a sub-per-mille precision relative to the transition energy, this result marks a milestone in the experimental exploration of SFQED.
%
\begin{table}[t]
\caption{Average electric field $\braket{\mathcal{E}}_{n\ell}$ at the mean orbital radius $\braket{r}_{n\ell}$ for selected kaonic-atom states, and its ratio $\chi$ to the Schwinger critical field 
$\mathcal{E}_c = m_e^2c^3/(e\hbar) \simeq 1.3 \times 10^{18}$~V/m.}
\label{tab:efield_schwinger}
\begin{ruledtabular}
\begin{tabular}{cccc}
State $n\ell$ & $\braket{r}_{n\ell}$~[fm] & $\braket{\mathcal{E}}_{n\ell}$~[V/m] & $\chi$\,=\,$\braket{\mathcal{E}}_{n\ell}/\mathcal{E}_c$ \\
\addlinespace[1mm]
\colrule
\addlinespace[1mm]
KF--3d &  66 & $4.91 \times 10^{18}$ & 3.70 \\
KF--4f & 113 & $1.48 \times 10^{18}$ & 1.11 \\
KF--5g & 172 & $0.59 \times 10^{18}$ & 0.44 \\
KF--6h & 244 & $0.28 \times 10^{18}$ & 0.21 \\
KF--7i & 328 & $0.15 \times 10^{18}$ & 0.06 \\
\end{tabular}
\end{ruledtabular}
\end{table}
Finally, to place KF in the broader strong-field context, Fig. \ref{fig:schwinger_lim} shows the average electric field $\braket{\mathcal{E}}_{n\ell}$ for circular kaonic states from KH to KAr. In KF, the ratio to the Schwinger critical field increases from $\chi$\,=\,0.44 for the 5g state to 3.70 for 3d, while higher-lying states remain below the critical value. This identifies a clear crossover into the strong-field regime as the kaon cascades from 5g to lower-$n$ states, with the 4f orbital already probing fields above the Schwinger scale. This feature is reflected in the size of the QED contributions: in KF the QED correction is already 67.54~eV for the 5g--4f transition and rises to 222.99~eV for the 4f--3d, showing the strong enhancement of BSQED effects at small orbital radii, where the field grows rapidly. The overall trend of the field in Fig. \ref{fig:schwinger_lim} follows qualitatively the hydrogenic scaling for circular states, $\braket{\mathcal{E}}_{n\ell}\propto Z^{3}/[n^{3}(n-1/2)]$ \cite{Bethe:1957ncq}, which explains both the rapid increase with atomic number $Z$ and the faster growth towards low-$n$ levels. In contrast to earlier studies that probed even larger values of $\chi$ in high-$Z$ exotic systems \cite{Powers:1980ie, Dubler:1978rq, gotta2008x}, the present work accesses transitions above the Schwinger limit that are free from strong-interaction effects and only weakly affected by other sources of uncertainty (see Table \ref{tab:KF_tab_SM}), enabling a direct and stringent test of BSQED in the strong-field regime.
%
\ifdraft
\section{Conclusions}
\fi

In summary, we have provided experimental evidence for SFQED effects in KF by measuring several low-$n$ x-ray transitions with SIDDHARTA-2 at DA$\Phi$NE. In the 10-50~keV range, we measure transitions such as 5g--4f, 6h--4f and 4f--3d, where the kaon experiences electric fields exceeding the QED critical value defined by the Schwinger limit, $\mathcal{E}_c = m_e^2c^3/(e\hbar) \simeq 1.3 \times 10^{18}$~V/m. The mean field surpasses the $\chi$\,=\,$\braket{\mathcal{E}}_{n\ell}/\mathcal{E}_c$\,=\,1 threshold already at the 4f level and reaches on the 3d state a ratio $\chi$\,=\,3.70. In this overcritical regime, QED corrections amount to as much as 222.99~eV for the 4f--3d transition, far exceeding estimated theoretical uncertainties associated with FNS, HFS, electronic screening and the kaon mass. The measured transition energies are in agreement with state-of-the-art \textsc{mcdfgme} calculations, including these sizable QED contributions. The 5g--4f x-ray transition -- the most intense measured line occurring above the Schwinger limit ($\chi$\,=\,1.11) -- shows an energy residual of 5.8\,$\pm$\,4.7\,(stat.)\,$\pm$\,5.5\,(syst.)~eV from the theoretical predictions and a $\sim$\,9$\sigma$ sensitivity to QED effects, providing a direct comparison between theory and experiment in the critical strong-field domain of QED. These results prove that x-ray transitions in kaonic atoms probe regimes above the QED critical field, confirming theory-experiment agreement in this strong-field regime. These results establish kaonic atoms as precision probes of BSQED in previously inaccessible strong-field regimes.
\newline

\textit{Acknowledgments}—We thank Jaroslava Obertova for estimating the strong interaction shifts and widths. We gratefully acknowledge the Polish high-performance computing infrastructure PLGrid (HPC Center: ACK Cyfronet AGH) for providing computer facilities and support within the computational grant no. PLG/2025/018524. We thank C. Capoccia from INFN-LNF and H. Schneider, L. Stohwasser, and D. Pristauz-Telsnigg from Stefan Meyer-Institut for their fundamental contribution in designing and building the SIDDHARTA-2 setup. We also thank INFN-LNF and the DA$\Phi$NE staff for the excellent working conditions and their ongoing support. Special thanks to Catia Milardi for her continued support and contribution during the data taking. Part of this work was supported by the INFN (KAONNIS project); the Austrian Science Fund (FWF): [P24756-N20 and P33037-N]; the Croatian Science Foundation under the project IP-2022-10-3878; the EU STRONG-2020 project (Grant Agreement No. 824093); the EU Horizon 2020 project under the MSCA (Grant Agreement 754496); the Japan Society for the Promotion of Science JSPS KAKENHI Grant No. JP18H05402, JP22H04917; the Polish Ministry of Science and Higher Education grant No. 7150/E-338/M/2018 and the Polish National Agency for Academic Exchange (grant no PPN/BIT/2021/1/00037). This article/publication is based upon work from COST ActionCA24131-ENRICH, supported by COST (European Cooperation in Science and Technology, http://www.cost.eu/).\newline

\textit{Data Availability}—Data and input files for the \textsc{mcdfgme} code are available upon reasonable request.
%
\bibliographystyle{apsrev4-2}
\bibliography{kf_prl}
\clearpage
\onecolumngrid
\begin{center}
    \textbf{\large Supplemental Material:\\ 
    Bound-state QED test beyond the Schwinger limit with kaonic fluorine 
    }
\end{center}
\vspace{0.5cm}
%
\section*{S1. Extended table for transition energies}
%
\renewcommand{\thefigure}{S\arabic{figure}}
\setcounter{figure}{0}
\renewcommand{\thetable}{S\arabic{table}}
\setcounter{table}{0}
\renewcommand{\theequation}{S\arabic{equation}}
\setcounter{equation}{0}
\begin{table*}[htb]
    \centering
    \caption{Measured and calculated transition energies for kaonic fluorine. Reported are the experimental energies $E_{if}^{(\mathrm{exp.})}$ with statistical ($\delta E_{if}^{(\mathrm{stat.})}$) and systematic ($\delta E_{if}^{(\mathrm{sys.})}$) uncertainties, compared with theoretical values $E_{if}^{(\mathrm{calc.})}$. Theoretical values include QED contribution ($E_{if}^{(\mathrm{QED})}$, split into first-order (QED1) and second-order (QED2) terms) and theoretical uncertainties are reported due to finite nuclear size (FNS), electron screening, the PDG kaon's mass, and hyperfine splitting (HFS). All values are given in eV.}
    \begin{ruledtabular}
    \begin{tabular}{cccccccccccc}
    \addlinespace[1mm]
    Transition & $E_{if}^{(\mathrm{exp.})}$ & $\delta E_{if}^{(\mathrm{stat.})}$ & $\delta E_{if}^{(\mathrm{sys.})}$ & $E_{if}^{(\mathrm{calc.})}$ & $E_{if}^{(\mathrm{QED})}$ & $E_{if}^{(\mathrm{QED1})}$ & $E_{if}^{(\mathrm{QED2})}$ & FNS & PDG & Screen. & HFS \\
    \addlinespace[1mm]
    \hline
    \addlinespace[1mm]
6h--5g & 12673.9 & 5.9 & 7.1 & 12684.69 & 23.49 & 23.31 & 0.17 & 0.01 & 0.33 & -0.09 & 0.27 \\
7i--5g & 20298.8 & 16.3 & 6.0 & 20327.49 & 32.26 & 32.02 & 0.24 & 0.01 & 0.52 & -0.25 & 0.27 \\
5g--4f & 23383.2 & 4.7 & 5.5 & 23377.38 & 67.54 & 67.07 & 0.47 & 0.03 & 0.60 & -0.04 & 0.87 \\
8k--5g & 25279.6 & 36.2 & 5.0 & 25285.56 & 35.65 & 35.38 & 0.27 & 0.01 & 0.65 & -0.48 & 0.27 \\
6h--4f & 36073.3 & 30.7 & 30.0 & 36062.07 & 91.02 & 90.38 & 0.64 & 0.04 & 0.93 & -0.13 & 0.87 \\
4f--3d & 50586.7 & 24.3 & 23.0 & 50590.48 & 222.99 & 221.45 & 1.54 & 0.13 & 1.30 & 0.01 & 4.16 \\
    \end{tabular}
    \end{ruledtabular}
    \label{tab:KF_tab_SM}
\end{table*}

\end{document}